\documentclass[pra,twocolumn,a4paper,showpacs,superscriptaddress]{revtex4}
\usepackage{amsmath}
\usepackage{amsfonts}
\usepackage{graphicx}
\begin{document}
\title{Constraints on the uncertainties of entangled symmetric qubits} 
\author{A. R. Usha Devi}
\email{arutth@rediffmail.com}
\author{M. S. Uma}
\author{ R. Prabhu}
\affiliation{Department of Physics, Bangalore University, Bangalore-560 056, India}
\author{A. K. Rajagopal}
\affiliation{ Department of Computer Science, George Mason University, Fairfax, Virginia, and
Inspire Institute Inc., McLean, Virginia.} 

\date{\today}

\begin{abstract}
We derive   necessary and sufficient inseparability conditions 
imposed on the variance matrix  of  symmetric qubits. These constraints are identified 
by examining  a structural parallelism between continuous variable states and 
two qubit states.  Pairwise entangled symmetric multiqubit states are  shown here 
to obey these  constraints. We also  bring out an elegant local invariant 
structure exhibited by our constraints. 
\end{abstract}


\maketitle

Entanglement is a central property of multipartite quantum systems as it forms 
the corner stone of all aspects of quantum information, communication, and computation
~\cite{zol}. The first task is to find a criterion if a given state 
is entangled  or not. Peres-Horodecki inseparability criterion~\cite{per-hor}\, viz., 
{\em positivity under partial transpose} (PPT)  has been extremely 
fruitful in addressing this question and provides  necessary and sufficient conditions for 
$2\times 2$ and $2\times3$ dimensional systems.   It is found that the PPT  criterion is significant 
in the case of infinite dimensional bipartite Continuous Variable (CV) states too. An important advance came about 
through an  identification of  how Peres-Horodecki criterion gets translated elegantly into the properties of 
the second moments (uncertainties) of CV states~\cite{simon}.  This results in  restrictions~\cite{simon, duan} 
on  the covariance matrix of an entangled bipartite CV state.   In the special case of  two-mode Gaussian states, 
where the basic entanglement properties are imbibed in the structure of  its covariance matrix,  the  restrictions on 
the covariance matrix are found~\cite{simon, duan} to be necessary and sufficient for inseparability. 
Investigations on the structure of variance matrix have proved to be  crucial in  understanding  the issue of 
entanglement in CV states~\cite{werner, gied1, gied2}  and 
a great deal of interest has been catching up in  experimentally  accessible, simple conditions of inseparability 
involving higher order moments~\cite{vog, aga, hil}.

In a parallel direction,  growing importance is being evinced towards quantum correlated
macroscopic atomic ensembles~\cite{hald, gere, madsen, juls1, juls2}. In the last few years 
experimental  generation of  entangled multiqubit states  in  trapped-ion 
systems,~\cite{Roos,Sackett,Leibfried,Haffner}
where individual particles can be manipulated, has been accomplished,  giving  new hopes for 
scalable quantum information processing. 

In the present work, we  explore a  structural parallelism in  CV states and  two qubit systems by constructing 
covariance matrix of the latter. We show that pairwise entanglement between any two qubits of a symmetric $N$ qubit 
state is completely characterised by the  off-diagonal block of the two qubit covariance matrix. We establish   
the inseparability  constraints satisfied by the covariance 
matrix and these are identified to be equivalent to the generalized spin squeezing inequalities~\cite{Korbicz} for 2 
qubit entanglement. The interplay  between  two basic principles viz, the uncertainty principle and the 
nonseparability gets highlighted through the restriction on the covariance matrix of a quantum correlated state. 

We first recapitulate succinctly  the approach employed by Simon~\cite{simon} for  bipartite CV states: 
The basic variables of bi-partite CV states are the conjugate quadratures of two field modes, $\hat{\xi}=(\hat{q}_1, 
\hat{p}_1, \hat{q}_2, \hat{p}_2)$,  which satisfy the canonical commutation relations
\begin{eqnarray}
\label{commu}
&&[\hat{\xi}_\alpha,\hat{\xi}_\beta]= i\,\Omega_{\alpha\beta} \ \ \ \ \ \ \alpha, \beta=1,2,3,4\, ; \nonumber \\
&& \Omega=\left(
\begin{array}{cc}
J & 0\\
0 & J
\end{array}
\right),\ \ \ \
J=\left(
\begin{array}{cc}
0 & 1\\
-1 & 0
\end{array}
\right). 
\end{eqnarray}
The $4 \otimes 4$ real symmetric covariance matrix of a bipartite CV state 
is defined through its elements \break $V_{\alpha\beta}=\frac{1}{2}\langle\{ 
\Delta\hat{\xi}_\alpha, \Delta\hat{\xi}_\beta \}\rangle,$ where
 $\Delta\hat{\xi}=\hat{\xi}_\alpha-\langle \hat{\xi}_\alpha\rangle$ and 
 $\{\Delta\hat{\xi}_\alpha, \Delta\hat{\xi}_\beta \}= 
 \Delta\hat{\xi}_\alpha\Delta\hat{\xi}_\beta+\Delta\hat{\xi}_\beta\Delta\hat
{\xi}_\alpha.$
Under canonical transformations, the variables of the two-mode system transform 
as $\hat{\xi}\rightarrow\hat{\xi'}=S\,\hat{\xi},$ where $S\in Sp\,(4,R)$ 
corresponds to a real symplectic $4\times 4$ matrix.
Under such transformations, the covariance matrix goes as $V~\rightarrow ~V'~=~ S\,V\,S^T$. 
The entanglement properties hidden in the covariance matrix $V$ remain unaltered under a 
local $Sp\,(2,R) \otimes Sp\,(2,R)$ transformation. It is convenient to caste the 
covariance matrix in a $2\times 2$ block form:
\begin{eqnarray}
\label{var}
V=\left(
\begin{array}{cc}
A & C\\
C^T & B
\end{array}
\right).
\end{eqnarray}
A local operation $S_1\oplus S_2\in Sp\,(2,R)\otimes  Sp\,(2,R)$ transforms  
the blocks $A,\ B,\ C$ of the variance matrix as
\begin{eqnarray}
\label{trans}
A\rightarrow A'=S_1\,A\,S^T_1,\ \ \ \ B\rightarrow B'=S_2\,B\,S^T_2,\nonumber \\
C\rightarrow C'=S_1\,C\,S^T_2.\hskip 0.5in
\end{eqnarray}
There are four local invariants associated with $V$:
$I_1={\rm det}\,A,\ I_2={\rm det}\,B,\ I_3={\rm det}\ C ,\  I_4={\rm Tr}\, (AJCBC^TJ)$
and the  Peres-Horodecki criterion imposes the restriction~\cite{simon}
\begin{eqnarray}
\label{cvinv} 
I_1\, I_2+\left(\frac{1}{4}-\vert I_3\vert\right)^2-I_4\geq\,\frac{1}{4}\,(I_1+I_2)
\end{eqnarray}
on the second moments of every  separable CV state. The signature of the invariant 
$I_3={\rm det}\,C$ has an important consequence: {\em Gaussian states with \,$I_3\geq 0$ 
are necessarily separable, whereas those with\, $I_3< 0$ and violating {\rm (\ref{cvinv})}
are entangled}. In other words, for Gaussian states violation of the 
condition (\ref{cvinv}) is both necessary and sufficient for entanglement.

Let us now turn our attention on qubits. An arbitrary two-qubit density operator belonging to  
the Hilbert-Schmidt space ${\cal H}={\cal C}^2\otimes{\cal C}^2$ is given by 
\begin{equation}
\rho=\frac{1}{4}\, [I\otimes I+\sum_{i=x,y,z} (\sigma_{1i}\, s_{1i}  
          +\sigma_{2i}\, s_{2i})
          +\sum_{i,j=x,y,z} \sigma_{1i}\, \sigma_{2j}\,t_{ij}\,]\, , 
\label{rho} 
\end{equation}
where $I$ denotes the $2\times 2$ unit matrix, $\sigma_{1\,i}=\sigma_i \otimes I,\, {\rm and} \ 
 \sigma_{2\,i}=I\otimes \, \sigma_i$\   
 ($\sigma_i$ are the standard Pauli spin matrices);   
$s_{\alpha\, i}={\rm Tr}\,(\rho\,  \sigma_{\alpha\, i}),\  \alpha=1,2,$  denote average spin components of   
 the $\alpha^{\rm th}$  qubit and  $t_{ij}={\rm Tr}\,(\rho\, \sigma_{1i}\sigma_{2j})$ 
are  elements of the real $3\times 3$ matrix  $T$ corresponding to two-qubit correlations. 
The set of 15 state parameters $\{  s_{1i},\  s_{2i},   \  t_{ij}\}$ transform~\cite{Hor}  under local unitary 
operations $U_1\otimes U_2$ on the qubits as, 
\begin{eqnarray}
\label{tran} 
s'_{1i}=\displaystyle\sum_{j=x,y,z} O^{(1)}_{ij}\, s_{1j},   &  s'_{2i}=
\displaystyle\sum_{j=x,y,z} O^{(2)}_{ij}\, s_{2j} \,,\nonumber\\
 t'_{ij}=\sum_{k,l=x,y,z} O^{(1)}_{ik}\, O_{jl}^{(2)}t_{kl} &  {\rm \ or \ }  
   \,\, T'=O^{(1)}\, T\,  O^{(2)\, T}\, ,  
    \end{eqnarray}    
where $O^{(\alpha)}\in SO(3)$ denote the  $3\times 3$ rotation matrices, 
corresponding uniquely to the $2\times 2$ unitary matrices $U_{\alpha}~\in~SU(2).$  

Our interest here is on two qubit states, which are completely symmetric under interchange. 
The symmetric two qubit states are confined to the three dimensional 
subspace $H_s={\rm Sym}\,({\cal C}^2\otimes{\cal C}^2)$ (`Sym' denotes 
symmetrization), spanned by the eigen states  $\left\{\ \vert  J_{\rm max}=1,\ M\rangle ;M=\pm 1,0 \right\}$  
of the total angular momentum of the qubits.  The state parameters of a symmetric two-qubit system obey the following   
constraints due to exchange symmetry~\cite{fn2}:
\begin{equation}
   \label{symcon}
s_{1i}= s_{2i}\equiv s_i\,  ,\ \  t_{ij}=t_{ji}\, ,\ \  {\rm Tr}\, (T)=1 \, , 
\end{equation} 
and  thus  8 real state parameters viz.,  $s_i$ and the elements $t_{ij}$ of the real symmetric correlation matrix 
$T$,  which has unit trace, characterize a symmetric two-qubit system. 

The basic variables of a two qubit system are  $\hat{\zeta}~=~\left(\sigma_{1i},\sigma_{2j}\right)$
and  the covariance matrix ${\cal V}$ of a two qubit system  may be  defined through
\begin{equation} 
\label{variance}
{\cal V}_{\alpha i;\, \beta  j}
=\frac{1}{2}\,\langle \{\Delta \hat{\zeta}_{\alpha\, i}, \Delta \hat{\zeta}_{\beta\, j} \}\rangle,
\  \alpha,\ \beta=1,2; \ i,j=x,\, y,\, z,
\end{equation} 
which can be written in the $3\times 3$ block form as
\begin{eqnarray}
\label{variance2}
{\cal V}=\left(
\begin{array}{cc}
{\cal A} & {\cal C}\\
{\cal C}^T & {\cal B}
\end{array}
\right),
\end{eqnarray}
where 
\begin{eqnarray}
\label{elem}
{\cal A}_{ij}&=&\frac{1}{2}\left[\langle\{\sigma_{1i},\sigma_{1j}\}\rangle
-\langle\sigma_{1i}\rangle\ \langle\sigma_{1j}\rangle\right] \nonumber \\
&=&\delta_{ij}- \langle\sigma_{1i}\rangle\ 
\langle\sigma_{1j}\rangle
=\delta_{ij}-s_{1i}s_{1j}, \nonumber\\
{\cal B}_{ij}&=&\frac{1}{2}\left[\langle\{\sigma_{2i},\sigma_{2j}\}\rangle
-\langle\sigma_{2i}\rangle\ \langle\sigma_{2j}\rangle\right]  \\ 
&=&\delta_{ij}- \langle\sigma_{2i}\rangle\ \langle\sigma_{2j}\rangle =\delta_{ij}-s_{2i}s_{2j},\nonumber\\
{\cal C}_{ij}&=&\frac{1}{2}\left[\langle{\sigma_{1i}\sigma_{2j}}\rangle
-\langle\sigma_{1i}\rangle\ \langle\sigma_{2j}\rangle\right]=t_{ij}-s_{1i}s_{2j},\nonumber \\
 {\rm or}& &  {\cal A}={\cal I}-s_1s_1^T,\  {\cal B}={\cal I}-s_2s_2^T, \   
 {\cal C}=T-s_1s_2^T.  \nonumber
\end{eqnarray}
Here  ${\cal I}$ denotes a $3\times 3$ identity matrix. 

In the case of symmetric states considerable simplicity ensues as a result of (\ref{symcon}) and  
we obtain  ${\cal A}~=~{\cal B}~=~1~-~ss^T,\ \ 
{\cal C}~=~T~-~ss^T.$    
We now establish an important property exhibited by  the 
off-diagonal block ${\cal C}$ of the covariance matrix ${\cal V}$ of  a symmetric two qubit state.

{\bf Lemma}: For every  separable symmetric state,  ${\cal C}~=~T-ss^T~$ is a positive 
definite matrix. 

{\bf Proof}: 
A two-qubit separable symmetric state is given by
\begin{equation}
\label{sep}
\varrho= \displaystyle\sum_w\, p_w\, \rho_{w}\otimes  
\rho_{w},\ \  
\displaystyle\sum_w p_w =1;\ \ 0\leq p_w\leq 1,
\end{equation}
where $\rho_{w}=\frac{1}{2}\,(1+\displaystyle\sum_{i=x,y,z}\, \sigma_i\ s_{iw} ),$  
denotes an arbitrary  single qubit density matrix.
The state variables $s_i$ and $t_{ij}$ \,of the two qubit separable symmetric  
system are given by 
\begin{eqnarray}
\label{sepst}
s_i&=&\langle\sigma_{\alpha\, i}\rangle={\rm Tr}\,\left(\varrho\, \sigma_{\alpha\, i}\right)
=\displaystyle\sum_w p_w\, s_{iw},\nonumber \\ 
t_{ij}&=&\langle\sigma_{1i}\,\sigma_{2j}\rangle={\rm Tr}\,\left(\varrho\, \sigma_ 
{1i} \sigma_{2j}\right)
=\displaystyle\sum_w p_w\, s_{iw}\, s_{jw}.\,\,\,\,\,\,
\end{eqnarray}
Let us now evaluate  the quadratic form $n^T(T-ss^T)\,n$ where $n\, (n^T)$ denotes any  
arbitrary real three componental  column (row),  
in a separable symmetric state:
\begin{eqnarray}
\label{proof}
n^T(T-ss^T)\,n=\sum_{i,j}\, (t_{ij}-s_i\,s_j)\, n_i\, n_j\hskip 1in  \nonumber\\
 =\sum_{i,j}\left[\displaystyle\sum_{w}  p_w\,s_{iw}\,s_{jw}-\sum_{w} p_w \,
s_{iw}\, \sum_{w^\prime} p_{w^\prime}\, 
s_{jw^\prime}\right]\, n_i\, n_j\nonumber\\
\hskip 0.2in =\sum_w p_w\, (\vec{s}\cdot\hat{n})^2-\left(\sum_w p_w\, 
(\vec{s}\cdot\hat{n})\right)^2 ,\hskip 0.8in  
\end{eqnarray}
which has the structure $\langle A^2\rangle - \langle A\rangle^2$ and is 
therefore a positive semi-definite quantity. 

The above lemma establishes the fact that the off diagonal block ${\cal C}$ of 
the covariance matrix is  {\em necessarily}   positive semidefinite for 
 separable symmetric states. And therefore, ${\cal C}<0$  serves as a sufficient 
 condition for inseparability in  two-qubit symmetric states.  

We investigate  pure entangled two qubit states.  A  Schmidt decomposed  pure 
entangled two-qubit state has the  form,
\begin{eqnarray}
\label{schm}
\vert\Phi\rangle=\kappa_1\,\vert \uparrow_1\, \uparrow_2\rangle+\kappa_2\,\vert 
\downarrow_1\,\downarrow_2\rangle,\hskip 0.2in\nonumber\\
 0<\kappa_2\, \leq \kappa_1<1,\ \  \kappa_1^2+\kappa_2^2=1.
\end{eqnarray}
(Here $\kappa_1,\ \kappa_2$ denote the  Schmidt coefficients). Obviously,  every 
pure entangled two-qubit state 
is   symmetric in the  Schmidt basis. It is easy to see that in (\ref{schm})
the $3\times 3$ correlation matrix $T$ is diagonal, 
\begin{equation}
\label{Tpure}
T={\rm diag}\, \left( 2\kappa_1 \kappa_2,\   -2\kappa_1 \kappa_2,\ 1\right)
\end{equation}
 and $s_i=\left(0,\, 0,\,\kappa_1^2-\kappa_2^2\right)$.  
The corresponding ${\cal C}$ matrix   also has  
a diagonal form 
\begin{equation}
\label{cpure}
{\cal C}=T-ss^T={\rm diag}\, \left(    
2\kappa_1 \kappa_2,\   -2\kappa_1 \kappa_2,\  4\, \kappa_1^2\,\kappa_2^2\right).
\end{equation}
It is readily  seen that ${\cal C}$ is  {\em negative},  as its 
diagonal element $-2\, \kappa_1\, \kappa_2<0,$ for an arbitrary entangled pure two-qubit  state. 
In other words, the condition ${\cal C}<0$ is both necessary and sufficient 
for pure entangled two-qubit states. 

This discussion of the entangled two-qubit pure state leads naturally to state 
the following theorem for the corresponding  symmetric  mixed state. 

{\bf Theorem:} The necessary condition for the inseparability of an arbitrary symmetric 
two qubit mixed state is given by ${\cal C}<0$.

{\bf Proof:} For the sake of brevity, we indicate here the steps leading to 
this condition and relegate the details to a separate communication~\cite{ARUnew}. 
An arbitrary two qubit symmetric state,  characterized by the density matrix  
(\ref{rho}),  with the state parameters obeying the permutation symmetry requirements 
(\ref{symcon}), 
gets transformed into a  $3\times 3$ block form, 
\begin{equation}
\label{symrho}
\rho_S=\scriptsize\frac{1}{4}\left(\begin{array}{ccc}1+2\, s_z+t_{zz} & 
a^*+b^*   & t_{xx}-t_{yy} 
 -2\, i\, t_{xy} \cr  
a +b  
   & 2\, (t_{xx}+t_{yy}) & 
a^*-b^* \cr 
t_{xx}-t_{yy}+2\, i\, t_{xy}  & a-b   & 1 - 2\, s_z+t_{zz} 
\end{array}  \right),   
\end{equation}
in the symmetric subspace characterized by the maximal value of 
total angular momentum $J_{\rm max}=1$ (with the ordering of the basis states given by $M=1, 0, -1$). Here, 
$a=\sqrt{2}\, (s_x+is_y),$ and $b=\sqrt{2}\, (t_{xz}+it_{yz}).$ 
The above $3\times 3$ matrix form  (\ref{symrho}) for $\rho_S$ is realized by a transformation from the two qubit 
basis to the total 
angular momentum basis $\vert J, M\rangle$ with $J=1, 0\, ; \, -J\leq M\leq J$: 
{\scriptsize\begin{eqnarray}
\label{symbasis}
\vert \uparrow_1\, \uparrow_2\rangle = \vert 1, 1\rangle, \ \  
\vert \downarrow_1\, \downarrow_2\rangle = \vert 1, -1\rangle \hskip 1in  &  \nonumber \\ 
 \vert \uparrow_1\, \downarrow_2\rangle =\frac{1}{\sqrt 2}( \vert 1, 0\rangle + \vert 0, 0\rangle ), \ \    
\vert \downarrow_1\, \uparrow_2\rangle = \frac{1}{\sqrt 2}( \vert 1, 0\rangle - \vert 0, 0\rangle ) &.  
\end{eqnarray}}
But, the partial transpose (PT)  of  (\ref{rho}), with respect to the second qubit  
  - identified as an operation leading to the complete sign reversal   
$\sigma_{2i}\rightarrow -\sigma_{2i}$  -  of a symmetric system   does not get restricted to the symmetric 
subspace with $J_{\rm max}=1$,  after the basis change (\ref{symbasis}). However, following the transformation 
(\ref{symbasis}) with another basis change 
{\scriptsize$\vert X\rangle ~=~\frac{-1}{\sqrt 2}(\vert 1,1\rangle ~-~ \vert 1, -1\rangle),\ 
\vert Y\rangle ~ =~\frac{-i}{\sqrt 2}(\vert 1,1\rangle ~+~ \vert 1, -1\rangle),\ 
\vert Z\rangle 
~=~\vert 1, 0\rangle $} leads to an elegant block structure for the
PT symmetric density matrix:  
\begin{equation}
\label{ptsym}
\displaystyle\rho^{T_2}_S=\frac{1}{2}\left(\begin{array}{cc} T & s \cr s^T & 1\end{array}   \right)
\end{equation}  
with $T$ being the $3\times 3$ two-qubit real symmetric correlation matrix and $s$ the 
$3\times 1$ column of qubit averages. As a final step, we identify the congruence 
\begin{equation}
\rho^{T_2}_S\sim \tilde\rho^{T_2}_S=L\, \rho^{T_2}_S\, L^\dag =
\frac{1}{2}\left(\begin{array}{cc} {\cal C} & 0 \cr 0 & 1
\end{array} \right),
\end{equation}
with $\scriptsize L=\left(\begin{array}{cc} {\cal I} & -s \cr 0 & 1 \end{array}\right),$ 
  leading us to the result   
\begin{equation} 
  \rho_S^{T_2} <0 \Leftrightarrow {\cal C}<0, 
  \end{equation}   
thus proving our theorem. $\Box$

We have therefore established that the ${\cal C}$ matrix approach 
 provides a simpler equivalent procedure to verify the inseparability status of a 
 symmetric two qubit system.

Now, we  proceed to explore how this basic structure ${\cal C}<0$ of inseparability reflects itself 
via   collective second moments of a symmetric $N$ qubit system. 
Collective observables of a $N$ qubit system are expressible in terms of the  angular momentum operator 
\begin{equation}
\label{ang} 
\vec{J}=\sum_{\alpha=1}^N\frac{1}{2}\,\vec{\sigma}_\alpha
\end{equation}
where $\vec{\sigma}_\alpha$ denote  the Pauli spin operator of the 
$\alpha^{\rm th}$ qubit. Symmetric $N$-qubit states are confined to the 
$N+1$ dimensional subspace $\left\{ |J_{\rm max}=N/2,\, M\rangle, \ -\frac{N}{2}\leq M\leq \frac{N}{2} \right\}$ of 
maximum angular momentum $J_{\rm max}=N/2$.  A collective  correlation matrix involving first and second moments of 
$\vec{J}$ may be defined by
\begin{equation} 
\label{vn} 
V^{(N)}_{ij}=\frac{1}{2}\langle J_i J_j+J_j J_i\rangle - \langle J_i\rangle \langle J_j \rangle \,\,;\,\,  i,\, j=x, 
y, z.
\end{equation}
The collective observables (upto second order in $\vec{J}$) in a  symmetric multiqubit state are expressible 
in terms of the constituent qubit variables as, 
\begin{eqnarray}
\label{mom}
\frac{1}{2}\langle (J_iJ_j+J_jJ_i)\rangle&=& \frac{1}{8}\,
\displaystyle\sum_{\alpha,\beta=1}^N\left\langle (\sigma_{\alpha
i}\sigma_{\beta j}
+\sigma_{\beta i}\sigma_{\alpha j})\right\rangle  \nonumber \\
&=&\frac{1}{4}\displaystyle
\sum_{\alpha,\beta=1}^N\left\langle (\sigma_{\alpha i}\sigma_{\beta
j})\right\rangle\nonumber \\
&=&\frac{N}{4}\,\left[ \delta_{i\, j}+ (N-1)\, t_{ij}\, \right], \nonumber \\ 
\langle J_i\rangle&=&\frac{1}{2}\sum_{\alpha=1}^N \langle \sigma_{\alpha
i}\rangle  =S_i=\frac{N}{2}\, s_i ,
\end{eqnarray}    
where we have made use of the following fact:    The bipartite  reductions $\rho_{\alpha\beta}$ of a symmetric 
multi-qubit state are all 
identical and  the average values of two qubit correlations 
$\langle(\sigma_{\alpha\,i}\sigma_{\beta\,j})\rangle=t_{ij}$ 
 - irrespective of the qubit labels $\alpha$ and $\beta$ - 
for any random pair of qubits drawn from a symmetric state. Moreover, $\langle\sigma_{\alpha\, i}\rangle=s_i$  
for all qubits belonging to a symmetric $N$-qubit system.

So, the correlation  matrix $V^{(N)}$ of (\ref{vn})  assumes the form 
\begin{equation}
\label{varmatrix}
V^{(N)}=\frac{N}{4}\left({\cal I}-ss^T+ (N-1)\, (T-ss^T)\right) 
\end{equation} 
 with ${\cal I}$ being a $3\times 3$ identity matrix\,; $T$ and $s$ are the state variables 
characterizing any two qubit partition  $\rho_{\alpha\beta}$ of a symmetric $N$ qubit system. 
It is convenient to express (\ref{varmatrix}) as 
\begin{equation}
\label{varmatrix2}
V^{(N)}+\frac{1}{N}\, SS^T=\frac{N}{4}\left({\cal I}+ (N-1)\, {\cal C}\right), 
\end{equation}
by shifting the second term i.e., $\frac{N}{4}\, ss^T$ to the left hand side and 
expressing $\frac{N}{2}\, s_i=\langle J_i\rangle=S_i$ (see (\ref{mom})), in terms of the collective average 
angular momentum. 
As has been established by  our Theorem, ${\cal C}$ is positive semi-definite for all separable symmetric two-qubit 
states,  implying that the  condition 
\begin{equation} 
\label{con}
V^{(N)}+\frac{1}{N}\, SS^T < \frac{N}{4}\, {\cal I}
\end{equation}
can only be satisfied by an entangled symmetric $N$ qubit  state. 

Under identical local unitary transformations 
~$U\otimes~ U\otimes ~\ldots ~\otimes~ U$~ on the qubits the variance matrix $V^{(N)}$ and the average spin $S$ 
transform as 
\begin{equation} 
\label{tran2}
V^{(N)'}=O\, V^{(N)}\, O^T\   {\rm and}\ S'=O\, S, 
\end{equation}
 where $O$ is a real orthogonal rotation matrix corresponding  to 
the  local unitary transformation $U$. Thus, the $3\times3$ real symmetric matrix $V^{(N)}+\frac{1}{N}\, SS^T$ can 
always be diagonalized by a suitable  identical local unitary transformation on all the qubits. In other words, 
 (\ref{con}) is a local invariant condition and it essentially implies:
  
{\em The symmetric $N$ qubit system is pairwise entangled iff the least eigen value of the 
real symmetric matrix ~$~V^{(N)}~+~\frac{1}{N}~\,~ SS^T$~
 is less than $N/4$,  }

{\bf Local invariant structure}:  Now, we  explore  how the negativity of the matrix 
${\cal C}$ reflects itself on the structure of the local invariants associated with the two qubit state~\cite{aru, 
aru2}. 

We denote the eigenvalues~\cite{fn} of the off-diagonal block  ${\cal C}$ of the covariance matrix (\ref{variance2}) 
by $c_1,\ c_2,\, {\rm and} \  c_3$.  Restricting   ourselves  to identical 
local unitary transformations~\cite{aru, aru2}, we define  three local invariants, which 
completely determine the eigenvalues of ${\cal C}$: 
\begin{eqnarray}
\label{newinv}
{\cal I}_1&=&\det\, ({\cal C})=c_1\,c_2\,c_3, \nonumber \\ 
 {\cal I}_2&=&{\rm Tr}\, ({\cal C})=c_1+c_2+c_3,\nonumber \\
  {\cal I}_3&=&{\rm Tr}\, ({\cal C}^2)=c_1^2+c_2^2+c_3^2. 
\end{eqnarray}  
The invariant  ${\cal I}_2$ may be rewritten as ${\cal I}_2~=~
{\rm Tr}\,(T~-~s\,s^T)~=~1-s_0^2$, since ${\rm Tr}\,(T)~=~1$ for a symmetric state~\cite{fn2}  
and we have denoted ${\rm Tr}\, (s\,s^T)=s_1^2+s_2^2+s_3^2=s_0^2$. Another useful invariant, 
which is a combination of the invariants defined through (\ref{newinv}),  may be 
constructed as 
\begin{equation}   
\label{aninv}
 {\cal I}_4=\frac{{\cal I}_2\,^2-{\cal I}_3}{2}=c_1\,c_2+c_2\,c_3+c_1\,c_3. 
 \end{equation}
 Positivity of the single qubit reduced density operator demands   
 $s_0^2\leq 1$  and leads in turn to the observation that the invariant 
 ${\cal I}_2$ is  positive for all symmetric states. Thus, all the three eigen values 
 $c_1,\, c_2,\, c_3$  of ${\cal C}$ can never  assume negative values for 
 symmetric qubits  and at most two of them can be negative.

We consider three distinct cases encompassing all pairwise  entangled symmetric 
states. 

Case (i): Let one of the eigenvalues $c_1=0$ and of the remaining two, let $c_2<0$ and $c_3>0$~\cite{neg}.

Clearly, the invariant ${\cal I}_1=0$ in this case. 
But we have 
\begin{eqnarray}
\label{cond2}
{\cal I}_4=c_2\, c_3 < 0,
\end{eqnarray}
which leads to a local invariant condition for two-qubit entanglement.

\noindent Case (ii): Suppose any two eigenvalues say, $c_1, c_2$, are  negative  and the 
third one $c_3$ is positive. 

Obviously, ${\cal I}_1>0$ in this case. But the invariant ${\cal I}_4$ 
assumes negative value:
\begin{eqnarray}
\label{cond3}
{\cal I}_4=c_1\, {\cal I}_2-c_1^2+c_2\, c_3<0 \\ \nonumber
\end{eqnarray}
as each term in the right hand side  is negative. In other words,  ${\cal I}_4<0$ 
gives the criterion for bipartite entanglement in this case too.

\noindent Case (iii): Let  $c_1<0$;   $c_2\  {\rm and}\  c_3$ be positive.

In this case we have 
\begin{eqnarray}
\label{cond1}
{\cal I}_1<0,  
\end{eqnarray}
giving the inseparability  criterion in terms of a local invariant.

In conclusion, we have investigated a structural similarity between continuous variable systems and 
symmetric two qubit states by constructing two-qubit variance matrix. 
We have shown here that the off-diagonal block of the variance matrix  ${\cal C}$ 
of a separable symmetric two qubit state is a positive semidefinite quantity.  
Symmetric two-qubit states satisfying the condition  
${\cal C}<0$ are therefore identified as inseparable. 
An equivalence between the Peres-Horodecki 
criterion and the negativity of the covariance matrix ${\cal C}$ is  
established, showing that our condition is both necessary  and sufficient 
for  entanglement in symmetric two qubit states.   
We have identified the constraints satisfied  by the collective correlation 
matrix $V^{(N)}$ of pairwise entangled symmetric N qubit states. 
Local invariant structure of our inseparability constraints is also investigated.  
In  a recent publication~\cite{Korbicz}, which appeared since 
 completion of this work, Korbicz et. al. have generalized the concept of spin squeezing 
 connecting it to  the theory of entanglement witnesses and proposed the inequality   
$\frac{4\langle\Delta J_n^2\rangle}{N}<1-\frac{4\langle J_n\rangle^2}{N^2},$
(where $J_n=\vec{J}\cdot\hat{n},$  $\hat{n}$ denoting an arbitrary unit vector, 
$\Delta J_n^2=\langle J_n^2\rangle-\langle J_n\rangle^2$)  
 as a necessary condition for two-qubit entanglement in symmetric states. 
After some simple algebra, we find that  this  inequality 
is equivalent to  ${\cal C}<0$, thus establishing  a connection between 
the covariance matrix and the theory of entanglement witnesses. 
Further, the approach outlined in this paper has been extended recently~\cite{cmatrix}  
to obtain constraints on higher order covariance matrices,  
 which in turn lead to a family of inseparability conditions for  
 various  even partitions of symmetric $N$-qubit  systems.    
 
We thank the Referees for their insightful comments which have 
improved the presentation of our work in this revised form.
  


\begin{thebibliography}{1}
\bibitem{fn2} Permutation symmetry demands that $\langle \sigma_{1 i}\rangle~=~\langle \sigma_{2 i}\rangle$ and 
$\langle\sigma_{1 i}\sigma_{2 j}\rangle~=~ \langle\sigma_{1 j}\sigma_{2 i}\rangle .$ 
Moreover, the squared total angular momentum operator of symmetric two qubits  
satisfies the condition $\langle(J_x^2~+~J_y^2~+ ~ J_z^2)\rangle~=~J_{\rm max}(J_{\rm 
max}~+~1)~=~\frac{N}{2}~(\frac{N}{2}~+~1)~=~2$ i.e.,   $\frac{1}{4}\langle(\vec{\sigma}_1+\vec{\sigma}_2)^2\rangle=2$. 
This leads, in  turn,  
to the unit trace condition ${\rm Tr}\, (T)=1$ on the 
correlation matrix of the two-qubit symmetric states.
\bibitem{fn} The off-diagonal block ${\cal C}$ of the covariance matrix can be diagonalized by 
an appropriate identical local unitary transformation on the qubits.
\bibitem{neg} We must note that if one of the eigenvalues i.e., $c_1,\, c_2,\, {\rm\ or\ } c_3$ 
is  equal to zero and another negative, the  remaining one  should be positive  
in order to preserve the condition\, ${\cal I}_2>0.$
\end{thebibliography}
\end{document}